\documentclass[10pt,journal,epsfig]{IEEEtran}
\usepackage[dvips]{graphicx}
\usepackage{graphicx}
\usepackage{amssymb}
\usepackage{cite}
\usepackage{subfigure}
\usepackage{multirow}
\usepackage{slashbox}
\usepackage{stfloats}
\usepackage{amsmath}

\usepackage{subeqnarray}
\usepackage{algorithm}
\usepackage{algpseudocode}
\usepackage{cases}
\usepackage[Symbol]{upgreek}
\makeatletter

\newcommand{\Rmnum}[1]{\expandafter\@slowromancap\romannumeral #1@}
\makeatother

\newtheorem{lemma}{Lemma}

\newtheorem{claim}{Claim}

\begin{document}
\IEEEoverridecommandlockouts
\title{GSVD-Based Precoding in MIMO Systems with Integrated Services}
\author{Weidong Mei, Zhi Chen, \IEEEmembership{Member, IEEE} and Jun Fang \IEEEmembership{Member, IEEE}
\thanks{Manuscript received June 27, 2016; revised August 15, 2016; accepted September 1, 2016. Date of publication; date of current version. This work was supported in part by the National Natural Science Foundation of China under Grant 61571089, and by the High-Tech Research and Development (863) Program of China under Grant 2015AA01A707. The associate editor coordinating the review of this manuscript and approving it for publication was Prof. Yong Xiang.}
\thanks{The authors are with National Key Laboratory of Science and Technology on Communications, University of Electronic Science and Technology of China, Chengdu (611731), China (e-mails: mwduestc@gmail.com; chenzhi@uestc.edu.cn; JunFang@uestc.edu.cn)
}}
\maketitle

\begin{abstract}
This letter considers a two-receiver multiple-input multiple-output (MIMO) Gaussian broadcast channel model with integrated services. Specifically, we combine two sorts of service messages, and serve them simultaneously: one multicast message intended for both receivers and one confidential message intended for only one receiver. The confidential message is kept perfectly secure from the unauthorized receiver. Due to the coupling of service messages, it is intractable to seek capacity-achieving transmit covariance matrices. Accordingly, we propose a suboptimal precoding scheme based on the generalized singular value decomposition (GSVD). The GSVD produces several virtual orthogonal subchannels between the transmitter and the receivers. Subchannel allocation and power allocation between multicast message and confidential message are jointly optimized to maximize the secrecy rate in this letter, subject to the quality of multicast service (QoMS) constraints. Since this problem is inherently complex, a difference-of-concave (DC) algorithm, together with an exhaustive search, is exploited to handle the power allocation and subchannel allocation, respectively. Numerical results are presented to illustrate the efficacy of our proposed strategies.
\end{abstract}
\begin{IEEEkeywords}
Physical-layer service integration (PHY-SI), GSVD, broadcast channel (BC), secrecy capacity region
\end{IEEEkeywords}

\section{Introduction}
\IEEEPARstart{H}{igh} transmission rate and secure communication are the basic demands for the future 5-Generation (5G) cellular networks. A heuristic way is to combine multiple coexisting services, e.g., multicast service and confidential service, into one integral service for one-time transmission, referred to as \emph{physical-layer service integration} (PHY-SI). Traditionally, service integration techniques rely on upper-layer protocols to allocate different services on different logical channels, which is quite inefficient. On the contrary, PHY-SI enables coexisting services to share the same resources by exploiting the physical characteristics of wireless channels, thereby significantly increasing the spectral efficiency.

The study of PHY-SI can be traced back to Csisz{\'a}r and K{\"o}rner's seminar work in \cite{csiszar1978broadcast}, where the fundamental limitation of PHY-SI is established in a discrete memoryless broadcast channel (DMBC). In recent years, this kind of approach
has gained renewed interest, especially that in various multi-antenna scenarios, such as Gaussian broadcast channels \cite{Hung2010Multiple, ekrem2012capacity}, and bidirectional relay channels \cite{Wyrembelski2012Physical}. Nonetheless, these works merely handle the PHY-SI from the viewpoint of information theory, i.e., derive capacity results or characterize coding strategies that result in certain rate regions \cite{Schaefer2014Physical}. As to how to design the transmit strategies to achieve these capacity regions, there are few works.

In this letter, we handle the PHY-SI from the view point of signal processing, i.e., design the precoding matrices of the transmitted messages. However, the resultant optimization problem is challenging to solve and it is physically difficult to eliminate the interference induced by the coupling of service messages. As a comprise, we pay our attention to some suboptimal but easy-to-implement transmit designs, e.g., the generalized singular value decomposition (GSVD). The basic merit of GSVD lies in its simplicity, since it yields several decoupled parallel subchannels between the transmitter and the receivers. In fact, GSVD-based precoding has been widely adopted in MIMO Gaussian broadcast channels for the purpose of confidentiality \cite{khisti2010secure2, fakoorian2012optimal, fakoorian2011dirty} or multicasting \cite{senaratne2013GSVD, senaratne2011beamforming, senaratne2010generalized}. It is natural to extend these results to the case with concurrent transmission of multicast message and confidential message.

This letter considers a two-receiver MIMO broadcast channel with two sorts of messages: a multicast message intended for both receivers, and a confidential message intended for merely one authorized receiver. The confidential message needs to be kept perfectly secure from the unauthorized receiver. Both messages are precoded by the matrices generated from GSVD. The resulting optimization problem turns out to be a secrecy rate maximization problem with quality of multicast service constraints, which is nonconvex by nature. To handle the nonconvexity, a difference-of-concave algorithm is proposed to determine the power allocation scheme for each subchannel. Based on the results, an exhaustive search is performed to determine the subchannel allocation scheme. By this means, the GSVD secrecy rate region could be derived.

\section{System Model}
We consider the downlink of a multiuser system in which a multi-antenna transmitter serves two receivers, and each receiver is equipped with multiple antennas. Both receivers have ordered the multicast service and receiver 1 (authorized receiver) further ordered the confidential service. The confidential message must be kept perfectly secure from receiver 2 (unauthorized receiver).

The received signal at receivers is modeled as
\begin{equation}\label{yk}
\begin{split}
  {{\mathbf{y}}_1} = \;{{\bf{H}}_1}{\bf{x}} + {{\mathbf{z}}_1},\;{{\mathbf{y}}_2} = \;{{\bf{H}}_2}{\bf{x}} + {{\mathbf{z}}_2},
\end{split}
\end{equation}
respectively, where ${{\mathbf{H}}_1}\in {{\mathbb{C}}^{{{N}_{b} \times {N}_{t}}}}$ (resp. ${{\mathbf{H}}_2}\in {{\mathbb{C}}^{{ {N}_{e} \times {N}_{t}}}}$) is the channel matrix from the transmitter to receiver 1 (resp. receiver 2), ${N}_{t}$, ${N}_{b}$ and ${N}_{e}$ are the number of antennas employed by the transmitter, receiver 1 and receiver 2, respectively. ${{\mathbf{z}}_1}$ and ${{\mathbf{z}}_2}$ are independent identically distributed (i.i.d.) complex Gaussian noise with zero mean and unit variance. ${{\mathbf{x}}}\in {{\mathbb{C}}^{{{N}_{t}}}}$ is the coded information, which consists of two independent components, i.e.,
\begin{equation}\label{x3c}
  {\bf{x\;}}\; = \;{{\bf{x}}_0} + \;{{\bf{x}}_c},
\end{equation}
where ${\bf{x}}_{0}$ is the multicast message intended for both receivers, and ${\bf{x}}_{c}$ is the confidential message intended only for receiver 1. We assume $\mathbf{x}_{0} \sim \mathcal{CN}(\mathbf{0},\mathbf{Q}_0)$, $\mathbf{x}_{c} \sim \mathcal{CN}(\mathbf{0},\mathbf{Q}_c)$ \cite{Hung2010Multiple}, where $\mathbf{Q}_0$ and $\mathbf{Q}_c$ are the transmit covariance matrices.

Denote $R_0$ and $R_c$ as the achievable rates associated with the multicast and confidential messages, respectively. Then the secrecy capacity region $C_s({\bf{H}}_1,{\bf{H}}_2,P)$ is given as the set of nonnegative rate pairs $(R_0,R_c)$ satisfying \cite{Hung2010Multiple}
\begin{subequations}\label{Region1}
\begin{align}
&{R_0} \le \mathop {\min }\limits_{k = 1,2} \log \left| {{\bf{I}} + {{\left( {{\bf{I}} + {{\bf{H}}_k}{{\bf{Q}}_c}{\bf{H}}_k^H} \right)}^{ - 1}}{{\bf{H}}_k}{{\bf{Q}}_0}{\bf{H}}_k^H} \right|,\\
&{R_c} \le \log \left| {{\bf{I}} + {{\bf{H}}_1}{{\bf{Q}}_c}{\bf{H}}_1^H} \right| - \log \left| {{\bf{I}} + {{\bf{H}}_2}{{\bf{Q}}_c}{\bf{H}}_2^H} \right|,
\end{align}
\end{subequations}
and $\text{Tr}(\mathbf{Q}_0+\mathbf{Q}_c) \le P$ with $P$ being the total transmit power budget at the transmitter.

With perfect CSI being available at the transmitter, to find capacity-achieving $\mathbf{Q}_0$ and $\mathbf{Q}_c$, we must solve the following secrecy rate maximization (SRM) optimization problem with quality of multicast service (QoMS) constraints, i.e.,
\begin{subequations}\label{op1}
\begin{align}
\nonumber g({r_{ms}}&)=\mathop {\max}\limits_{{{\bf{Q}}_0},{{\bf{Q}}_c}} \log \left| {{\bf{I}} + {{\bf{H}}_1}{{\bf{Q}}_c}{\bf{H}}_1^H} \right| - \log \left| {{\bf{I}} + {{\bf{H}}_2}{{\bf{Q}}_c}{\bf{H}}_2^H} \right|\\
\text{s.t.}\; &\log \left| {{\bf{I}} + {{\left( {{\bf{I}} + {{\bf{H}}_k}{{\bf{Q}}_c}{\bf{H}}_k^H} \right)}^{ - 1}}{{\bf{H}}_k}{{\bf{Q}}_0}{\bf{H}}_k^H} \right| \ge {r_{ms}} ,k=1,2 \label{op1a}\\
&\text{Tr}({{\bf{Q}}_0} + {{\bf{Q}}_c}) \le P,\\
&{{\bf{Q}}_0} \succeq {\bf{0}}, {{\bf{Q}}_c} \succeq {\bf{0}},
\end{align}
\end{subequations}
where $r_{ms}$ is predetermined requirement of the achievable multicast rate. To derive the boundary points of the secrecy capacity region $C_s({\bf{H}}_1,{\bf{H}}_2,P)$, one should traverse all possible ${r_{ms}}$'s and store the corresponding optimal objective value $g({r_{ms}})$, and then the rate pair $({r_{ms}},g({r_{ms}}))$ is a boundary point of the secrecy capacity region.

However, the coupling of confidential message and multicast message renders problem (\ref{op1}) nonconvex and thus intractable to solve. On the other hand, it makes the interference cancelation operation difficult for receivers. These facts motivate us to devise some simple but physically realizable alternatives. Naturally, the concept of GSVD-based precoding becomes attractive, since it can perfectly decouple all data streams.

\section{GSVD-Based Precoding Scheme for PHY-SI}
In this section, we will show our proposed GSVD-based precoding design for PHY-SI mathematically. First, let us introduce the GSVD via the following lemma.
\begin{lemma}(GSVD transform, \cite[Definition 1]{fakoorian2012optimal})\label{GSVD}
Given two matrices ${{\mathbf{H}}_1} \in {{\mathbb{C}}^{{{N}_{b} \times {N}_{t}}}}$ and ${{\mathbf{H}}_2}\in {{\mathbb{C}}^{{ {N}_{e} \times {N}_{t}}}}$, GSVD transform returns two unitary matrices ${{\bf{\Psi }}_r} \in {{\mathbb{C}}^{{{N}_{b} \times {N}_{b}}}}$ and ${{\bf{\Psi }}_e} \in {{\mathbb{C}}^{{{N}_{e} \times {N}_{e}}}}$, and non-negative diagonal matrices $\mathbf{C}$ and $\mathbf{D}$, and a matrix ${\bf{A}} \in {{\mathbb{C}}^{{{N}_{t} \times q}}}$ with $q=\min({N}_{t},{N}_{b}+{N}_{e})$, such that
\begin{equation}\label{gsvd}
\begin{split}
&{{\bf{H}}_1}{\bf{A}} = {{\bf{\Psi }}_r}{\bf{C}},\\
&{{\bf{H}}_2}{\bf{A}} = {{\bf{\Psi }}_e}{\bf{D}}.
\end{split}
\end{equation}
The nonzero elements of ${\bf{C}}$ are in ascending order while the nonzero elements of ${\bf{D}}$ are in descending order. Moreover, ${{\bf{C}}^T}{\bf{C}} + {{\bf{D}}^T}{\bf{D}} = {\bf{I}}$. Let $c_i$ and $d_i$ represent the $i^{\text{th}}$ diagonal elements of ${{\bf{C}}^T}{\bf{C}}$ and ${{\bf{D}}^T}{\bf{D}}$, respectively.
\end{lemma}

If matrices ${{\bf{H}}_1}$ and ${{\bf{H}}_2}$ represent the wireless channels as we have defined in (\ref{yk}), with the precoding matrices $\rho{\bf{A}}$ at the transmitter and receiver reconstruction matrices ${{\bf{\Psi }}_r^H}/\rho$ and ${{\bf{\Psi }}_e^H}/\rho$ at the respective receiver, we will get $q$ noninterfering broadcast subchannels between the transmitter and the receivers. The coefficient $\rho$ denotes transmit power normalization. The gains of those subchannels are determined by the diagonal elements of ${\bf{C}}$ and ${\bf{D}}$. Let us define the subchannels with condition $c_i=1$ (resp. $d_i=1$) as \emph{private channels} (PCs) of receiver 1 (resp. receiver 2), and the subchannels with condition $0 < c_i,d_i <1$ as \emph{common channels} (CCs) of both receivers.

The detailed number of CCs and PCs realized through GSVD under different system configurations is given in Table \ref{tab} as below. Herein we point out two cases under which the GSVD-based PHY-SI is infeasible. First, we know from Table \ref{tab} that when $N_t \ge N_b+N_e$, GSVD precoding will not generate any CCs. Thus, multicast message cannot be transmitted under this case. Second, if $c_i \le d_i$ holds for all $i$, the achieved secrecy rate would always be zero even without multicasting \cite[Claim 1]{khisti2010secure2}, which invalidates the confidential message transmission. In this letter, we will only focus on the nontrivial cases where GSVD-based PHY-SI is feasible.
\begin{table}[h]
  \centering
  \caption{Numbers of CCs and PCs Realized Through GSVD Precoding \cite{senaratne2013GSVD}}\label{tab}
    \begin{tabular}{lccc}
    \hline
    \multirow{2}[4]{*}{System Configuration} & \multirow{2}[4]{*}{\#CC} & \multicolumn{2}{c}{\#PC} \\
               &       & Receiver 1 & Receiver 2 \\
    \hline
    C1: $N_t < N_b$, $N_e \le N_t$ &  $N_e$ & $N_t - N_e$ & 0 \\
    C2: $N_t \ge N_b$, $N_e > N_t$ & $N_b$ & 0     & $N_t - N_b$ \\
    C3: $N_t \le N_b$, $N_e \ge N_t$ & $N_t$ & 0     & 0 \\
    C4: ${N_b} < {N_t},{N_e} < {N_t},$ & $N_b +N_e - N_t$ & $N_t -N_e$ & $N_t -N_b$ \\
    ${N_b} + {N_e} > {N_t}$ & & &\\
    C5: $N_b +N_e \le N_t$ & 0     & $N_b$ & $N_e$ \\
    \hline
    \end{tabular}%
\end{table}%

Next we define two sets ${\Gamma _0}$ and ${\Gamma _c}$ with cardinality $\left| {{\Gamma _0}} \right|=M$ and $\left| {{\Gamma _c}} \right|=N$, corresponding to the indices of subchannels allocated to the multicast message and confidential message, respectively. These two sets satisfy $\Gamma _0 \cup \Gamma _c = \{ 1,2,...,q\}$ and $\Gamma _0 \cap \Gamma _c = \emptyset $. Without loss of generality, we assume ${\Gamma _0}=\{ {i_1},{i_2},...,{i_M}\}$ and ${\Gamma _c}= \{ {j_1},{j_2},...,{j_N}\}  = \{ 1,2,...,q\} \backslash {\Gamma _0}$. Both receivers are assumed to be aware of ${\Gamma _0}$ and ${\Gamma _c}$.

Now, let the transmitted signal vectors ${\bf{x}}_{0}$ and ${\bf{x}}_{c}$ be constructed as
\begin{subequations}\label{precd}
\begin{align}
&{{\bf{x}}_0} = {{\bf{A}}_0}{{\bf{s}}_0},\;{{\bf{s}}_0} \sim \mathcal{CN}({\bf{0}},{{\bf{P}}_0})\\
&{{\bf{x}}_c} = {{\bf{A}}_c}{{\bf{s}}_c},\;{{\bf{s}}_c} \sim \mathcal{CN}({\bf{0}},{{\bf{P}}_c})
\end{align}
\end{subequations}
where ${{\bf{A}}_0} = {\bf{A}}{{\bf{E}}_0}$, ${{\bf{A}}_c} = {\bf{A}}{{\bf{E}}_c}$ with ${{\bf{E}}_0} \buildrel \Delta \over = \left[ {{{\bf{e}}_{{i_1}}},{{\bf{e}}_{{i_2}}},...,{{\bf{e}}_{{i_M}}}} \right]$ and ${{\bf{E}}_c} \buildrel \Delta \over = \left[ {{{\bf{e}}_{{j_1}}},{{\bf{e}}_{{j_2}}},...,{{\bf{e}}_{{j_N}}}} \right]$, ${\bf{e}}_l \in {\mathbb{R}}^{q}, l=1,2,...,q$ represents the $l$th column vector of ${\mathbf{I}}_{q}$, ${\bf{A}}$ is obtained from the GSVD transform in (\ref{gsvd}). ${{\bf{s}}_0}\in {{\mathbb{C}}^{M}}$ and ${{\bf{s}}_c\in {{\mathbb{C}}^{N}}}$ represent input data symbols of multicast message and confidential message, respectively. ${{\bf{P}}_0}$ and ${{\bf{P}}_c}$ are positive semi-definite diagonal matrices representing the power allocated by the transmitter to the data symbols ${{\bf{s}}_0}$ and ${{\bf{s}}_c}$, respectively. Substituting (\ref{precd}) into the channel model (\ref{yk}) and using (\ref{gsvd}) yields
\begin{subequations}\label{yk2}
\begin{align}
&{{\bf{y}}_1} = {{\bf{\Psi }}_r}{\bf{C}}{{\bf{E}}_0}{{\bf{s}}_0}+{{\bf{\Psi }}_r}{\bf{C}}{{\bf{E}}_c}{{\bf{s}}_c}+{{\bf{z}}_1},\\
&{{\bf{y}}_2} = {{\bf{\Psi }}_e}{\bf{D}}{{\bf{E}}_0}{{\bf{s}}_0}+{{\bf{\Psi }}_e}{\bf{D}}{{\bf{E}}_c}{{\bf{s}}_c}+{{\bf{z}}_2}.
\end{align}
\end{subequations}
Consequently, the secrecy rate $R_c$ in (\ref{Region1}) can be expressed as
\begin{align}
\nonumber{R_c} &= \log \frac{{\left| {{\bf{I}} + {{\bf{\Psi }}_r}{\bf{C}}{{\bf{E}}_c}{{\bf{P}}_c}{\bf{E}}_c^H{{\bf{C}}^H}{\bf{\Psi }}_r^H} \right|}}{{\left| {{\bf{I}} + {{\bf{\Psi }}_e}{\bf{D}}{{\bf{E}}_c}{{\bf{P}}_c}{\bf{E}}_c^H{{\bf{D}}^H}{\bf{\Psi }}_e^H} \right|}}\\
\nonumber&\mathop  = \limits^{(a)} \log \left| {{\bf{I}} + {{\bf{P}}_c}{{({\bf{C}}{{\bf{E}}_c})}^H}{\bf{C}}{{\bf{E}}_c}} \right| - \log \left| {{\bf{I}} + {{\bf{P}}_c}{{({\bf{D}}{{\bf{E}}_c})}^H}{\bf{D}}{{\bf{E}}_c}} \right|\\
&\mathop  = \limits^{(b)} \sum\limits_{n = 1}^N {\log (1 + {p_{c,n}}c_{{j_n}}^2)}  - \sum\limits_{n = 1}^N {\log (1 + {p_{c,n}}d_{{j_n}}^2)},\label{sec}
\end{align}
where ${p_{c,n}}$ is the $n$th diagonal element of ${{\bf{P}}_c}$, equality ($a$) is due to the fact that ${\bf{\Psi }}_r^H{{\bf{\Psi }}_r} = {\bf{\Psi }}_e^H{{\bf{\Psi }}_e} = {\bf{I}}$ and the Sylvester's determinant theorem \cite{harville1997matrix}, i.e., $\det ({\bf{I}} + {\bf{UV}}) = \det ({\bf{I}} + {\bf{VU}})$ for appropriate dimensions of $\bf{U}$ and $\bf{V}$, and equality ($b$) is due to the fact that any two columns of ${\bf{C}}$ (or ${\bf{D}}$) are orthogonal.

Since all subchannels are parallel and ideally non-interfering, the multicast message transmission is able to experience a clean link without the interference of confidential message. Thus, in the same way as (\ref{sec}), the achievable multicast rate w.r.t. receiver 1 and receiver 2 is given by
\begin{equation}\label{multi}
{R_{0,1}} = \sum\limits_{m = 1}^M {\log (1 + {p_{0,m}}c_{{i_m}}^2)}, \; {R_{0,2}} = \sum\limits_{m = 1}^M {\log (1 + {p_{0,m}}d_{{i_m}}^2)},
\end{equation}
respectively, where ${p_{0,m}}$ is the $m$th diagonal element of ${{\bf{P}}_0}$.

The transmit power allocated to multicast message and confidential message is therefore determined by
\begin{equation}\label{totPower}
\begin{split}
{\rm{Tr}}({{\bf{x}}_0}{\bf{x}}_0^H) &= {\rm{Tr}}({{\bf{A}}_0}{{\bf{P}}_0}{\bf{A}}_0^H)= \sum\nolimits_{m = 1}^M {{a_{0,m}}{p_{0,m}}},\\
{\rm{Tr}}({{\bf{x}}_c}{\bf{x}}_c^H) &= {\rm{Tr}}({{\bf{A}}_c}{{\bf{P}}_c}{\bf{A}}_c^H)= \sum\nolimits_{n = 1}^N {{a_{c,n}}{p_{c,n}}},
\end{split}
\end{equation}
where ${a_{0,m}}$ is $m$th diagonal element of ${\bf{A}}_0^H{{\bf{A}}_0}$, and ${a_{c,n}}$ is $n$th diagonal element of ${\bf{A}}_c^H{{\bf{A}}_c}$. Hence, the resultant QoMS-constrained SRM problem is given by
\begin{subequations}\label{op2}
\begin{align}
\nonumber&\mathop {\max }\limits_{\left\{p_{c,n}\right\}_{n = 1}^N,{\Gamma _c}\atop\left\{p_{0,m}\right\}_{m = 1}^M,{\Gamma _0}} \sum\limits_{n = 1}^N {\log (1 + {p_{c,n}}c_{{j_n}}^2)} - \sum\limits_{n = 1}^N {\log (1 + {p_{c,n}}d_{{j_n}}^2)} \\
\text{s.t.}\;&\sum\nolimits_{m = 1}^M {\log (1 + {p_{0,m}}c_{{i_m}}^2)} \ge {r_{ms}},\label{op2a}\\
&\sum\nolimits_{m = 1}^M {\log (1 + {p_{0,m}}d_{{i_m}}^2)}  \ge {r_{ms}},\label{op2b}\\
&\sum\nolimits_{m = 1}^M {{a_{0,m}}{p_{0,m}}}+\sum\nolimits_{n = 1}^N {{a_{c,n}}{p_{c,n}}} \le P,\label{op2c}\\
&{p_{c,n}} \ge 0, {p_{0,m}} \ge 0, \forall n,m,\label{op2d}\\
&{\Gamma _0}=\{ {i_1},{i_2},...,{i_M}\}, {\Gamma _c}= \{ {j_1},{j_2},...,{j_N}\}.
\end{align}
\end{subequations}

\section{A Tractable Approach to the SRM problem}
Problem (\ref{op2}) decouples the confidential message and multicast message; however, it couples the subchannel allocation and power allocation to each subchannel. To solve (\ref{op2}), our strategy is to determine the power allocation scheme with a given subchannel allocation scheme. Then by exhausting all possible subchannel allocation schemes, we could find the maximum secrecy rate.

\subsection{Subchannel Allocation Scheme}
Although we resort to the exhaustive search to handle the subchannel allocation, the following criterions could help us reduce the computational complexity.

\begin{claim}\label{UAPC}
The PCs of unauthorized receiver should be discarded, for they cannot transmit either confidential message or multicast message.
\end{claim}
\begin{claim}\label{APC}
The PCs of authorized receiver can only be used for the confidential message transmission.
\end{claim}
\begin{claim}\label{CC}
For any CC satisfying $c_i \le d_i$, it can only be used for the multicast message transmission.
\end{claim}
\begin{IEEEproof}
For Claim \ref{UAPC} and \ref{APC}, it is easy to see that the PCs must be invalid to the multicast message transmission, since only one receiver is able to receive the multicast message. Claim \ref{CC} could be verified by contradiction. Assume that the maximum secrecy rate can be achieved when CCs satisfying $c_i \le d_i$ are used for confidential message transmission, then a larger secrecy rate will always be attained if we specify the power allocated to these subchannels as zero, which is contrary to the assumption.
\end{IEEEproof}

As a result, when performing the exhaustive search, we can limit our searching scope to CCs with condition $c_i > d_i$.

\subsection{Power Allocation Scheme}
With a given subchannel allocation scheme, we need to solve the following optimization problem, i.e.,
\begin{equation}\label{op3}
\begin{split}
\mathop {\max }\limits_{\{p_{c,n}\}_{n = 1}^N,\atop\{p_{0,m}\}_{m = 1}^M} &\sum\limits_{n = 1}^N {\log (1 + {p_{c,n}}c_{{j_n}}^2)} - \sum\limits_{n = 1}^N {\log (1 + {p_{c,n}}d_{{j_n}}^2)} \\
\text{s.t.}\; &\text{(\ref{op2a})-(\ref{op2d}) satisfied}.
\end{split}
\end{equation}
However, problem (\ref{op3}) remains nonconvex because of its objective function. Moreover, due to the additional QoMS constraints, it is difficult to obtain its closed-form solutions by directly checking its Karush-Kuhn-Tucker (KKT) conditions, as \cite{fakoorian2012optimal,fakoorian2011dirty} did. To handle it, we propose a difference-of-concave (DC) approach to (\ref{op3}). Its basic idea is to locally linearize the nonconcave part $-\sum\nolimits_{n = 1}^N {\log (1 + {p_{c,n}}d_{{j_n}}^2)}$ at some feasible point $\{p_{c,n}^{(i)}\}_{n = 1}^N$ via first-order Taylor expansion and iteratively solve the linearized problem, i.e.,
\begin{equation}\label{op4}
\begin{split}
(\{ p_{c,n}^{(i+1)}\} _{n = 1}^N,\{ p_{0,m}^{(i+1)}\} _{m = 1}^M&) \in \\
\mathop {\max }\limits_{\{p_{c,n}\}_{n = 1}^N,\{p_{0,m}\}_{m = 1}^M} &g(\{ {p_{c,n}}\} _{n = 1}^N;\{ p_{c,n}^{(i)}\} _{n = 1}^N)\\
\text{s.t.}\; &\text{(\ref{op2a})-(\ref{op2d}) satisfied},
\end{split}
\end{equation}
where $g(\{ {p_{c,n}}\} _{n = 1}^N;\{ p_{c,n}^{(i)}\} _{n = 1}^N) \buildrel \Delta \over = \sum\nolimits_{n = 1}^N {\log (1 + {p_{c,n}}c_{{j_n}}^2)}  - \sum\nolimits_{n = 1}^N {\log (1 + p_{c,n}^{(i)}d_{{j_n}}^2)}  - \sum\nolimits_{n = 1}^N {\frac{{d_{j,n}^2({p_{c,n}} - p_{c,n}^{(i)})}}{{\ln 2(1 + p_{c,n}^{(i)}d_{j,n}^2)}}}$.

Problem (\ref{op4}) is a convex problem, which can be optimally solved by interior-point method (IPM) \cite{boyd2009convex}. Denoting the objective secrecy rate returned by $i$th iteration as $R^i$, as a basic property of DC \cite{NIPS2009}, we immediately have the following conclusion: Every limit point of $\{R^i\}_i$ is a stationary point of problem (\ref{op3}).

We summarize the above-developed DC approach to problem (\ref{op3}), together with the exhaustive search over subchannel allocation schemes, in Algorithm 1. Notice that in line \ref{AlgUpt} of Algorithm 1, we diminish the size of $\Phi$ by directly eliminating the infeasible subchannel allocation scheme.
\begin{algorithm}
  \caption{Subchannel and power allocation strategies for solving (\ref{op2})}
  \begin{algorithmic}[1]
    \State Initiate $r_{ms}=0$, $\delta > 0$ and $\epsilon > 0$, and let $\Phi$ be the collection of all possible subchannel allocation schemes;
    \State \textbf{while $\Phi  \ne \emptyset$ do}
    \State \quad Let $k=1$, $\left|\Phi\right|=K$ and $\Phi  = \{ {\Phi _1},{\Phi _2}, \cdots ,{\Phi _K}\}$;
    \State \quad \textbf{while $k \le K$ do}
    \State \qquad Fix ${\Gamma _0}$ and ${\Gamma _c}$ by allocating subchannels to different service messages according to $\Phi_k$;
    \State \qquad Set $i=0$, $R^{k,i}=0$ and $\{p_{c,n}^{(k,i)}\}_{n = 1}^N$ such that $\sum\nolimits_{n = 1}^N {{a_{c,n}}{p_{c,n}^{(k,i)}}} \le P$;
    \State \qquad \textbf{Repeat}
    \State \qquad\quad $i=i+1$;
    \State \qquad\quad Solve problem (\ref{op4}) via IPM and get $p_{c,n}^{(k,i)}$;
    \State \qquad\quad \textbf{if} problem (\ref{op4}) is infeasible \textbf{then}
    \State \qquad\qquad $R^{k,i}=0$, $\Phi=\Phi  - \{\Phi_k\}$;\label{AlgUpt}
    \State \qquad\qquad \textbf{jump to} line \ref{AlgRpt};
    \State \qquad\quad \textbf{end if}
    \State \qquad\quad Compute $R^{k,i}=\sum\nolimits_{n = 1}^N {\log (1 + {p_{c,n}^{(k,i)}}c_{{j_n}}^2)}  - \sum\nolimits_{n = 1}^N {\log (1 + {p_{c,n}^{(k,i)}}d_{{j_n}}^2)}$;
    \State \qquad \textbf{Until $\left| {{R^{k,i}} - {R^{k,i-1}}} \right| < \epsilon $}
    \State \qquad $R^{k}=R^{k,i}$\label{AlgRpt}
    \State \qquad $k=k+1$
    \State \quad \textbf{end while}
    \State \quad Let $R({r_{ms}}) = \arg \mathop {\max }\limits_{k = 1,2,...,K} {R^k}$, and store the rate pair $\left( {{r_{ms}},R({r_{ms}})} \right)$;
    \State \quad Update ${r_{ms}}={r_{ms}}+\delta$;
    \State \textbf{end while}
  \end{algorithmic}
\end{algorithm}

\section{Numerical Results}
In this section, we provide numerical results to illustrate the secrecy rate region derived from our proposed GSVD-based scheme, compared with the secrecy capacity region obtained from exhaustive search over the set $\left\{ {\left. {\left( {{{\bf{Q}}_c},{{\bf{Q}}_0}} \right)} \right|} {{\bf{Q}}_0} \succeq {\bf{0}}, {{\bf{Q}}_c} \succeq {\bf{0}}, \text{Tr}({{\bf{Q}}_0} + {{\bf{Q}}_c}) \le P \right\}$, and the traditional time division multiple address (TDMA)-based service integration strategy, which assigns the confidential message and multicast message to two different orthogonal time slots. For the fairness of comparison, the secrecy rate and multicast rate achieved by TDMA should be \textbf{halved} \cite{Wyrembelski2012Physical}.

In the simulation, the channels are randomly generated from i.i.d. complex Gaussian distribution with zero mean and unit variance. The number of antennas at the transmitter, authorized receiver and unauthorized receiver are $N_t=3$, $N_b=4$ and $N_e=3$, corresponding to C1 and C3 in Table \ref{tab}, and the transmit power $P$ is set as $20$dB.

\begin{figure}[!t]
\centering
\includegraphics[width=2.6in]{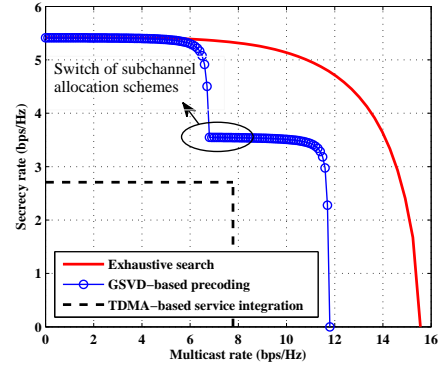}
\DeclareGraphicsExtensions.
\caption{Secrecy rate regions achieved by different schemes}\label{SRR1}
\vspace*{-12pt}
\end{figure}
Fig.\,\ref{SRR1} plots the secrecy rate region achieved by different strategies. The secrecy capacity region serves as a reference indicating the performance loss the GSVD-based scheme would inevitably experience. One can notice that there exist a switching point at the boundary of the GSVD secrecy rate region. Actually, it is caused by the switch of different subchannel allocation schemes. From Fig.\,\ref{SRR1}, we find that at low QoMS region, the GSVD-based scheme achieves identical performance to the secrecy capacity region. This is attributed to the near-optimality of GSVD-based precoding at high signal-to-noise ratio (SNR) in the confidential message transmission \cite{khisti2010secure2}. However, with the increase of QoMS, the gap between these two regions gradually expands. This performance degradation is due to the suboptimality of GSVD-based precoding to the multicast message transmission. Even so, our proposed scheme achieves significantly larger rate region than the TDMA-based one. Note that the motivations to use GSVD and TDMA are both to decouple the confidential message and multicast message. It follows that GSVD-based decoupling gives better performance than TDMA, which implies the inherent advantage of PHY-SI over traditional service integration.

In addition, we examined by simulations that our observations above also apply to C2 and C4 in Table \ref{tab}. The results are not shown here due to the page limit. This observation suggests that it is sound to use GSVD-based precoding for achieving service integration at PHY.

\section{Conclusion}
In this letter, we consider a GSVD-based precoding design for two-receiver MIMO broadcast channel with PHY-SI. The GSVD precoding matrices of confidential message and multicast message are designed to maximize the achievable secrecy rate while satisfying the QoMS constraints. Since this QoMS-constrained SRM problem is simultaneously associated with the optimization of subchannel allocation and power allocation, we combine an exhaustive search over subchannel allocation schemes with a DC algorithm to solve it. Numerical results show that the GSVD-based scheme outperforms the traditional TDMA-based service integration and attains the boundary of secrecy capacity region at low QoMS region.

\bibliography{PHYSI_GSVD}
\bibliographystyle{IEEEtran}

\end{document}